\documentstyle[epsf]{l-aa}
\input psfig.tex

\begin{document}

   \thesaurus{08         % Stars
              (08.14.1;  % Stars: neutron
               02.05.2;  % Physical data and processes: Equation of state
               02.08.1;  % Physical data and processes: Hydrodynamics
               02.09.1;  % Physical data and processes: Instabilities
               02.14.1)} % Physical data and processes: Nuclear reactions, 
                         % nucleosynthesis, abundances
%
   \title{The fate of a neutron star just below the minimum mass: 
does it explode?}

%   \subtitle{Does it explode?}

   \author{K. Sumiyoshi    \inst{1,2}  \and
           S. Yamada       \inst{1,3}  \and
           H. Suzuki       \inst{1,4}  \and
           W. Hillebrandt  \inst{1}
          }

   \offprints{K. Sumiyoshi}

   \institute{Max-Planck-Institut f\"ur Astrophysik,
              Karl-Schwarzshild-Str. 1, 85740 Garching, Germany\
              \and
              The Institute of Physical and Chemical Research (RIKEN), 
              Hirosawa, Wako, Saitama 351-01, Japan\
              \and
              Department of Physics, School of Science, 
              The University of Tokyo, Hongo, Bunkyo, Tokyo 113, Japan\
              \and
              High Energy Accelerator Research Organization (KEK), 
              Oho, Tsukuba, Ibaraki 305, Japan\
             }
   \date{Received July dd, 1997; accepted mm dd, 1997}

   \maketitle

   \begin{abstract}
First results of numerical simulations are presented which compute
the dynamical evolution of a neutron star with a mass slightly below
the minimum stable mass by means of a new implicit (general relativistic)
hydrodynamic code. We show that such a star first undergoes a 
phase of quasi-static expansion, caused by slow nuclear $\beta$-decays,
lasting for about 20 seconds, but then explodes violently.
The kinetic energy of the
explosion is around $10^{49} erg$, the peak luminosity
in electron anti-neutrinos is of order $10^{52} erg/s$, and the thermodynamic
conditions of the expanding matter are favorable for r-process
nucleosynthesis. These results are obtained for the Harrison-Wheeler
equation of state and a simple and, possibly, unrealistic 
treatment of $\beta$-decay rates and nuclear fission, which were
adopted for comparison with previous works.
However, we do not expect that the outcome will change
qualitatively if more recent nuclear input physics is used.
 
Although our study does not rely on a specific scenario of
how a neutron star starting from a bigger (and stable) mass
can reach the dynamical phase, we implicitly assume that the
final mass-loss event happens on a very short time scale, i.e.,
on a time scale shorter than a sound-crossing time, by removing
a certain amount of mass as an initial perturbation.
This assumption implies that the star has no time to adjust
its nuclear composition to the new mass through a sequence 
of quasi-equilibria. In the latter case, however, there exists no 
stable configuration below the minimum mass, because the 
equation of state of fully catalyzed matter is too soft.
Therefore, the dynamics of the explosion 
will not be too different from what we have obtained 
if different initial perturbations are assumed.

   \end{abstract}

%
%  14.Sep.'90: Demo-Vs.
%________________________________________________________________

\section{Introduction}
Ever since the discovery of binary neutron star systems
their final fate has been subject of intensive work and also of
many speculations. There are no doubts that because of  
orbital momentum loss by gravity waves the neutron stars
will ultimately merge together, ending as a fast spinning
black hole or massive neutron star, depending on the initial total
mass and the amount of mass possibly lost from the
system during the merger event. To the outside world
such a merger will manifest itself by a burst of gravitational
radiation, probably the best candidate for gravitational
wave detectors, a burst of neutrinos and, likely, also
a  burst of $\gamma$-rays. It has also been speculated 
that during the merger a small amount of neutron star matter
may be ejected which could, in principle, have nuclear
abundances similar to those commonly attributed to the r-process,
posing interesting limits on the rate of neutron star 
mergers over the history of the Galaxy.
An alternative scenario which motivates in part our present
work is the speculation
that one of the neutron stars
(the one with initially less mass) might lose most of
its mass to the companion prior to the merging event.
It may then end with a mass near the minimum
equilibrium configuration. Such a scenario requires
conservative mass-transfer, certain properties of the
neutron star's equation of state, etc., and it may be difficult to find the
right conditions. On the other hand, it cannot be excluded on simple grounds. 
 
Most investigations so far start from the assumption
that both neutron stars have nearly equal masses, an assumption 
which is in fact supported by the observed binary neutron star
systems with well determined masses. Also, nearly all neutron
stars seem to have masses at least consistent with
a gravitating mass of about 1.4 $M_{\odot}$. The final coalescence
phase of two neutron stars of initially 1.5 $M_{\odot}$ has
recently been studied extensively by Ruffert et al. (1996, 1997).
They performed 2- and 3-dimensional simulations with 
state-of-the-art micro-physics, but Newtonian hydrodynamics.
The outcome of their simulations was a final black hole,
a strong burst of neutrinos, accompanied by $\gamma$-rays
(with insufficient luminosity to explain the observed
$\gamma$-ray bursts at cosmological distance), and ejection 
of some heavy and neutron-rich elements. So in case of (nearly)
equal masses of both neutron stars it is very unlikely that
one of them (i.e. the one with the slightly lower initial
mass) will undergo slow mass-loss by Roche-lobe overflow
and end up near the minimum mass prior to merging.
 
However, the situation may be different for neutron star binaries with
very different initial masses. In that case the Roche-model 
may be more applicable and a scenario outlined by Imshennik and Popov
(1996) may result, namely the smaller neutron star may, due to its considerably 
larger radius, lose some fraction of its mass before dynamic mass transfer
sets in. Whether or not this can happen will depend on the true equation
of state of neutron star matter, on the mass-ratio of the two stars, 
on whether or not conservative mass transfer can be obtained for certain times,
etc. Investigating these questions is beyond the aim of this paper,
and a definite answer is hard to get (see however, e.g., Lai et al. 1994).  
It might well happen that in a few rare cases a neutron star
pair is born with largely different initial 
masses and one of them  might reach the critical (minimum) mass.
If it should explode at this stage this would have several interesting 
observable consequences (see, e.g., Eichler et al. (1989)). For example, the
more massive component could obtain a kick-velocity of up to 2000 $km/s$
(Imshennik and Popov, 1996), well in the range of observed pulsar
proper motions (Lyne and Lorimer, 1994). They would be the source of considerable
emission of $\gamma$-rays and neutrinos, and they should eject matter
of rather unusual chemical composition.
A second possibility to form neutron stars near the minimum mass, which
we just mention in passing, could be by fragmentation 
of rapidly rotating cores of collapsing massive stars.
 
One can, of course, turn  the argument around and ask: Would the 
impact of such explosions on galactic nucleosynthesis be so strange
that we can exclude them immediately? Or can we impose so strong constraints
on their event rate that they would be of little interest?
It is this latter argument which motivated our work and which led us to
reinvestigate the numerical studies of Colpi et al. (1989. 1991, 1993)
with improved numerical techniques (this paper) and with more realistic
micro-physics input data (forth-coming papers). 
Consequently, our final aim is to make 
firm predictions for the yields of neutron-rich isotopes ejected during the 
explosion of a neutron star just below the minimum mass, but the aim of the 
present paper less ambitious. We only want to demonstrate that
even under unfavorable conditions,
e.g. for an equation of state that predicts the transition to homogeneous
nuclear matter at rather low density (the Harrison-Wheeler equation of
state) neutron stars do explode if their mass drops below the minimum
mass allowed by an equation of state of matter in
nuclear statistical {\it and} $\beta$-equilibrium. This result is in qualitative
agreement with that obtained by Colpi et al. (1993), 
but it is more rigorous for the following reason.
The early evolution of such a neutron star and the transition
from a slow expansion to the fast explosion proceeds on
the time scale of (slow) $\beta$-decays, and its
modeling requires a code that is able to follow the evolution
over many dynamical times. Colpi et al. (1993) circumvented this problem
by rescaling the $\beta$-decay rates in order to make
both time scales comparable, an approach that is not fully satisfactory.
Here, we did not have to use this trick because we could compute
the evolution by means of an {\it implicit} hydro-code which is
described in Section 2. In order to be able to make a detailed comparison 
with their work we left the micro-physics input essentially unchanged
(see Section 2).
 
Given this assumption, e.g., the zero temperature Harrison-Wheeler equation of state 
and $\beta$-equilibrium, the minimum possible mass of a neutron star is
$M_{min}~=~0.189 M_{\odot}$ and it has a central density of $\rho _c~=~
2.67 \times 10^{13} g/cm^3$. It has a radius of $230 km$ and consists  
of a core region of homogeneous nuclear matter of about $0.13 M_{\odot}$
extending to $18 km$ and a crust that makes up for the rest. 
This initial configuration is identical to the one used by Colpi et al. (1993).
We then take off some mass from the surface and follow the evolution
by means of our implicit hydro-code. The results are given in Section 3,
including first estimates of the nucleosynthesis yields that can be
expected from the explosion,
and a discussion and summary conclude the paper. 

\section{The Hydrodynamical Model}
\subsection{Hydrodynamics}
We compute the evolution of a neutron star below the minimum mass 
by solving the equations of general relativistic hydrodynamics in 
spherical symmetry.  
The equations are solved numerically by an implicit 
method using  Lagrange (baryon mass) coordinates.  The code was 
developed and has been tested for a series of basic problems 
by  Yamada (1997).
In the present application we use a mesh of 100 equal mass-zones and
micro-physics input, such as 
the equation of state, nuclear reactions and neutrino cooling, 
is implemented in a way 
described below into the original 
hydrodynamics code.  

In addition,
we have modified the original 
code to handle time-dependent nuclear reactions and neutrino cooling.  
In order to guarantee stability, an 
entropy equation was added to 
the implicit hydrodynamics scheme (Yamada 1997b),
in addition to the energy equation. 
The temperature is calculated from the entropy,
and the internal energies obtained from the energy equation 
and the equation of state are used to check 
for numerical errors only.  
The entropy equation reads (Meyer 1989), 
\begin{equation}
T \frac{dS}{dt} = Q_{beta} + Q_{fission} - Q_{cooling}, 
\label{eqn:entropy}
\end{equation}
where S is the entropy per baryon, 
$Q_{beta}$ and $Q_{fission}$ are the heating rate per baryon due to 
the energy released by $\beta$-decays and fission of nuclei, 
respectively.  
$Q_{cooling}$ is the cooling rate per baryon due to neutrino emission 
through $\beta$-decays.  
The equation for the evolution of the electron fraction, 
Eq. (\ref{eqn:ye}), 
is not solved implicitly but is
followed separately by an explicit scheme.  
We first update the value of the electron fraction 
by the Runge-Kutta method and then solve the hydrodynamics 
with the new values of the electron fraction.  
The time step is constrained by limiting variations of all 
quantities to be less than 10 $\%$.  

We point out that in contrasts to earlier studies (e.g. Colpi et al (1993))
we adopted an implicit hydro code rather than an explicit method.
Therefore, time steps are not
constrained by the Courant condition as in explicit schemes.
Because of this advantage 
we can study the 
evolution of a neutron star just below the minimum mass for  
time scales as long as several seconds or more, which is necessary 
because the early evolution is governed
by the time scales of $\beta$-decays which are much longer than,
e.g., the sound crossing time which would limit the time step
in an explicit code.

\subsection{The Equation of State}
One of the main aims of the present paper is to demonstrate 
that neutron stars with a mass slightly less than the minimum
mass encounter an instability and explode. Since we want to compare
our numerical results, based on a newly developed hydrodynamics code,
with the so far most elaborate study of Colpi et al. (1993) we decided
to adopt the same equation of state as they did, namely the Harrison-Wheeler
equation of state (Harrison et al. 1965; see also Shapiro $\&$ 
Teukolsky, 1983) for neutron star matter in $\beta$-equilibrium, and 
an extension of it for matter out of $\beta$-equilibrium 
and at finite temperatures (Colpi et al. 1989).  
Details of this treatment are given below. One has to keep in mind,
however, that the Harrison-Wheeler equation of state is not very
realistic at densities near nuclear matter density and that it predicts
a transition to homogeneous nuclear matter far below the saturation density.
Consequently, for more realistic equations of state, neutron stars
near the minimum mass will have no central core of homogeneous nuclear matter
but will consist mainly of crust material. This will lead to more 
violent explosions than we find in our study which thus, in a sense,
is conservative if we want to answer the question whether or not these
objects explode at all. In a subsequent paper we will discuss the results 
obtained for more realistic equations of state.

\subsubsection{The Equilibrium Equation of State}
According to the Harrison-Wheeler equation of state, cold neutron star matter 
has three different phases: homogeneous nuclear matter and electrons
above the transition density 
$\rho_{tr} \equiv 4.3 \times 10^{12} g/cm^{3}$, a mixture of neutrons,
nuclei, and electrons down to the neutron-drip 
density $\rho_{dr} \equiv 3.2 \times 10^{11} g/cm^{3}$, and nuclei  
and electrons below  $\rho_{dr}$.
The highest density regime will be referred to as the
neutron star's {\it core}.  The density range $\rho_{tr} 
\geq \rho \geq \rho_{dr}$ 
is called the {\it inner crust}, and the remaining part  
is the {\it outer crust}.  
We recall again that the composition of the central part 
of the neutron star depends on the equation of state, and  
that for neutron stars near the minimum mass 
nuclei exist even in the central part (Colpi et al. 1993) 
when microscopically better motivated equations of state such 
as the BPS equation of state (Baym et al. 1971) 
are adopted. 

The properties of cold neutron star matter are determined by 
several equilibrium conditions at zero temperature.  
$\beta$-equilibrium with vanishing neutrino chemical potential 
is maintained in the whole star and determines the electron 
fraction $Y_{e} \equiv \frac{n_{e}}{n_{B}}$ together with the condition 
of charge neutrality.  Nuclear species in the inner and outer 
crust are determined so as to minimize the total energy per 
nucleon of the matter in $\beta$-equilibrium and  
charge neutrality.  The equilibrium between neutrons inside 
nuclei and dripped neutrons is maintained additionally 
in the inner crust.  
These conditions determine the composition of the matter, i.e. 
the electron fraction $Y_{e}$, the number fraction of nuclei 
$Y_{A} \equiv \frac{n_{A}}{n_{B}}$, the mass number $A$ and 
proton number $Z$ of the nuclei.  Other quantities such as the pressure 
are calculated accordingly with the composition.  
We treat neutrons, protons and electrons in the core and 
dripped neutrons and electrons in the crust as ideal Fermi gases.  
Properties of nuclei in the crust are calculated by means 
of the mass formula of the Harrison-Wheeler model.  
We use this equilibrium
equation of state of cold neutron star matter 
to construct the initial model 
(see section \ref{chap:ini}).

\subsubsection{The Non-Equilibrium Equation of State}
The equilibrium conditions described above are usually assumed if 
cold neutron stars in hydrostatic equilibrium are studied.  In order 
to investigate the instability of neutron stars below the minimum 
mass, however, one has to take into account departures from equilibrium 
because nuclear reaction times can be considerably longer 
than the hydrodynamical time scale which is of order milliseconds.  

At low temperatures
$\beta$-decays of free neutrons in the neutron star's core
are suppressed due to the limited phase space available in the highly degenerate 
environment there (Colpi et al. 1989).  
Therefore, the composition of 
the core is frozen and the electron fraction, $Y_{e}$, 
is fixed to its initial value during the evolution.  

The $\beta$-decay times of nuclei in the crust of a neutron star  
range over many orders of magnitude and play an essential 
role in initiating the dynamical instability.  These time scales 
are determined by the properties of the matter out of 
$\beta$-equilibrium and will be described in more detail in section 
\ref{chap:beta}. Due to the decay of nuclei
the electron fraction in the crust 
changes from its initial value 
once the matter moves out of $\beta$-equilibrium.  
The change of the electron fraction leads to a change of 
the pressure and affects the hydrodynamics, which in turn 
changes the thermodynamic conditions which determine 
the $\beta$-decay rates. 

Neutrons inside and outside the nuclei in the 
inner crust remain in equilibrium during the evolution since 
the density of free neutrons and the temperature
are high enough for the 
time scales of neutron captures (and their inverse
process) to be much shorter than the $\beta$-decays 
in the expanding neutron star matter (Meyer 1989).  
Therefore, we can assume that the equilibrium between neutrons inside and 
outside the nuclei is maintained during the evolution.  
We stress that 
the combination of rapid neutron captures with slow $\beta$-
decays resembles the classical conditions for r-process nucleosynthesis.  

In a neutron stars crust a complete statistical equilibrium
is not achieved
unless the temperature becomes very high (above $10^{10} K$; see, e,g.,
Lattimer et al. (1985)).  
The neutron star matter is heated by the energy 
released by the $\beta$-decay and fission of nuclei 
during the expansion of the neutron star.  
The amount of heating depends on details 
of the hydrodynamical evolution and on micro-physics data
such as the equation of 
state and the $\beta$-decay rates,  
and only if the temperature becomes sufficiently high 
one has to compute the equation of state in nuclear statistical equilibrium 
including weak, electro-
magnetic and strong interactions.  
In the present study, 
we assume one species of nuclei instead of computing the equilibrium
composition for a network of nuclei.
Moreover, we assume that
the number fraction of nuclei, $Y_{A}$, is fixed 
during the evolution because charged particle reactions are essentially frozen.  
For calculating the equation of state,
this latter assumption can be justified by the fact that
the temperatures of neutron star matter we find
in the current study never exceed a few times 10$^9$K. 

\subsubsection{Properties of the Expanding Matter}
Since we calculate the hydrodynamics in a Lagrangian mesh  
and can ignore transport effects,
we follow the evolution of each fluid mass element independently 
and evaluate the equation of state and the composition 
at every time step accordingly.
To be more precise, the properties of expanding neutron star matter
are obtained in the following way.  

The equation of state of the matter at high densities, 
$\rho \geq \rho_{tr}$, is calculated for a mixture of 
ideal Fermi gases of neutrons, protons and electrons.  
The composition in each mass element is determined by 
the electron fraction and charge neutrality.  
The electron fraction, $Y_{e}(j,t)$, 
of the Lagrangian mass element indexed by $j$ at time $t$ 
is fixed to its initial value 
\begin{equation}
Y_{e}(j,t) = Y_{e}(j,0) = Y_{e}^{0}(j). 
\end{equation}
 
If the mass element originally belonged 
to the core of the initial model and its density reaches $\rho_{tr}$ 
due to expansion, 
we assume that the matter is converted into a mixture of 
nuclei and ideal Fermi gases of neutrons and electrons.  
The composition ($Y_{e}$ and $Y_{A}$) is assumed to be 
that of cold neutron star matter at the transition density, 
$\rho_{tr}$.  
The entropy is assumed to be continuous over the phase transition.  
Afterwards, such a mass element evolves like an element composed of 
crust material as will be described below.  

The equation of state at densities 
$\rho \leq \rho_{tr}$ is calculated as a mixture of 
nuclei and ideal Fermi gases of neutrons and electrons.  
The composition is determined by the electron fraction, 
the number fraction of nuclei, the equilibrium 
condition for the neutrons and baryon number conservation.  
The electron fraction evolves because of  $\beta$-decays 
according to
\begin{equation}
\dot{Y_{e}}(j,t) = \lambda_{\beta}(j,t) Y_{A}(j,t), 
\label{eqn:ye}
\end{equation}
where $\lambda_{\beta}(j,t)$ is the $\beta$-decay rate 
of the nuclei in the mass-zone $j$ at time $t$.  The initial 
condition for Eq. (\ref{eqn:ye}) is the value of 
the electron fraction in the initial model.  
For mass elements originally belonging 
to the core 
equation (\ref{eqn:ye}) is solved only for times 
after the density reaches $\rho_{tr}$, and the initial value
of $Y_{e}$ is taken to be the equilibrium value at $\rho_{tr}$.  
The number fraction of nuclei, $Y_{A}(j,t)$, is kept constant, 
\begin{equation}
Y_{A}(j,t) = Y_{A}(j,0) = Y_{A}^{0}(j), 
\label{eqn:ya}
\end{equation}
unless after $\beta$-decay the spontaneous fission line is reached. 
For mass elements originally in the core, the equilibrium 
value of $Y_{A}$ at $\rho_{tr}$ is adopted as the value of 
$Y_{A}^{0}$ in Eq. (\ref{eqn:ya}).
The proton number of nuclei is calculated from the relation 
\begin{equation}
Z(j,t) = \frac{Y_{e}(j,t)}{Y_{A}(j,t)}.  
\label{eqn:z}
\end{equation}
The chemical potentials of neutrons inside and outside 
nuclei are balanced, 
\begin{equation}
\mu_{n}(j,t) \equiv \mu_{n}^{in}(j,t) = \mu_{n}^{out}(j,t).  
\label{eqn:mun}
\end{equation}
Here, $\mu_{n}^{in}(j,t)$ is the neutron chemical potential 
inside the nuclei and is calculated as a function of proton 
number, $Z$, and mass number, $A$, of nuclei obtained from the mass 
formula used in the Harrison-Wheeler equation of state.  
The chemical potential of dripped neutrons, $\mu_{n}^{out}(j,t)$, 
is given by the ideal Fermi gas expression.  
The condition of the baryon number conservation is expressed as 
\begin{equation}
A(j,t) = \frac{1}{Y_{A}(j,t)} (1-Y_{n}(j,t)), 
\label{eqn:a}
\end{equation}
where $Y_{n}$ is the number fraction of dripped neutrons 
defined by $Y_{n} \equiv \frac{n_{n}}{n_{B}}$.  
The density of dripped neutrons, $n_{n}$, is calculated 
in terms of the chemical potential, $\mu_{n}^{out}$, and 
the temperature, $T$.  

The mass number of nuclei, $A$, and the neutron chemical 
potential, $\mu_{n}$, are determined by solving the equations for
the chemical equilibrium of the neutrons, Eq. (\ref{eqn:mun}), and 
baryon number conservation, Eq. (\ref{eqn:a}).  
Numerically, we search for a solution $Y_{n}$ in the range 
between 0 and 1 by comparing the values of $\mu_{n}^{out}$ 
and $\mu_{n}^{in}$.  
$\mu_{n}^{out}$ is determined through $Y_{n}$ and 
$\mu_{n}^{in}$ is determined through $A$ and $Z$, which 
are given by Eqs. (\ref{eqn:z}) and (\ref{eqn:a}), respectively.  

When the matter is not in equilibrium and/or at finite 
temperature, the transition density from inner crust to outer 
crust is not necessarily in accord with the neutron-drip 
density $\rho_{dr}$ for the cold neutron star matter.  
Whether the matter belongs to the inner or outer crust 
is solely determined by the free neutron fraction, $Y_{n}$.  
If $\mu_{n}^{in} \leq \mu_{n}^{out}$ at $Y_{n} = 0$, 
nuclei are bound and there are no dripped neutrons. 
Therefore, the matter belongs to the outer crust.  
In this case, the equation of state is 
calculated as a mixture of nuclei and an ideal Fermi gas of 
electrons.  The composition is determined by the electron 
fraction and the number fraction of nuclei as in Eqs. 
(\ref{eqn:ye}) and (\ref{eqn:ya}).  The nuclear species are 
determined from Eqs. (\ref{eqn:z}) and 
(\ref{eqn:a}) with $Y_{n} = 0$.  

In order to take into account temperature effects, 
we evaluate all quantities of the Fermi gases of neutrons, 
protons,  and electrons at finite temperature.  
We use the mass formula in the Harrison-Wheeler equation of 
state even at finite temperature.  This should be a good 
approximation for temperatures well below $10^{10} K$.  
The entropy of the nuclei is approximated by that of a  non-degenerate 
ideal Fermi gas.  

\subsection{$\beta$-decay Rates}\label{chap:beta}
The $\beta$-decay rates appropriate for neutron star matter are 
evaluated in the same manner as was done by Colpi et al. (1989).  
The $\beta$-decay of free neutrons is completely suppressed 
in the core and in the inner crust due to the high electron degeneracy
there.  Therefore, only nuclei in the crust can decay.

The $\beta$-decay rate of nuclei, $\lambda_{\beta}(j,t)$, can be estimated
(Lattimer et al. 1977)
\begin{equation}
\lambda_{\beta} = \langle \frac{\rho}{ft} \rangle
( \frac{\Delta^{6}}{6} + \Delta^{5}\mu_{e} 
+ \frac{5}{2} \Delta^{4} \mu_{e}^{2} ),
\label{eqn:lamda}
\end{equation}
where $\mu_{e}$ is the chemical 
potential of electrons and $\Delta$ is defined by 
\begin{equation}
\Delta \equiv \mu_{n} - \mu_{p} - \mu_{e}.  
\end{equation}
This quantity measures the deviation from  $\beta$-equilibrium
and is equal to zero when $\beta$-equilibrium
is achieved.  
It is equal to the energy available for the decay, 
given by the $Q_{\beta}$ value of the parent nucleus minus 
the electron Fermi energy, $\mu_{e}$.  
The coefficient $\langle \frac{\rho}{ft} \rangle$ 
contains nuclear structure information (matrix elements
and level density in the daughter nuclei) and is rather
uncertain. 
Here we take this parameter to be $10^{-2.8} MeV^{-6}s^{-1}$ 
following the work of Colpi et al. (1993).  
We note that the value of this coefficient 
has an ambiguity ranging from $10^{-5.5}$ to $10^{-2.8} 
MeV^{-6}s^{-1}$, when $\Delta$ and $\mu_{e}$ are 
measured in units of MeV (Lattimer et al. 1977).  
The energy release by $\beta$-decays enters 
the entropy equation, Eq. (\ref{eqn:entropy}), and 
is expressed as (Meyer 1989), 
\begin{equation}
Q_{beta} = - \sum_{i} \mu_{i} \frac{dY_{i}}{dt} 
= \Delta \frac{dY_{e}}{dt}, 
\end{equation}
where the sum is over neutrons, 
protons and electrons, and 
we have used the charge neutrality condition.  

\subsection{Neutrino Cooling}
In the $\beta$-decays of 
nuclei  anti-neutrinos are emitted.
It is assumed that they escape freely from the 
star and contribute only by cooling the matter inside 
the star (Meyer 1989).  This is a fair approximation since
the time scale for the expansion of the star is much longer
than neutrino diffusion times.
The average energy of anti-neutrinos 
is evaluated by
\begin{equation}
\varepsilon_{\bar{\nu}_{e}} = \frac{3}{7} \Delta 
\frac{\Delta^{2} + 7 \Delta \mu_{e} + 21 \mu_{e}^{2}} 
     {\Delta^{2} + 6 \Delta \mu_{e} + 15 \mu_{e}^{2}}, 
\end{equation}
in accord with the $\beta$-decay rates adopted here (Lattimer et al. 1977).  
The local luminosity of anti-neutrinos per baryon is then calculated from
\begin{equation}
l_{\bar{\nu}_{e}} = 
\varepsilon_{\bar{\nu}_{e}} Y_{A} \lambda_{\beta}.  
\end{equation}
Finally, the total neutrino luminosity is obtained by integrating over 
the whole star, 
\begin{equation}
L_{\bar{\nu}_{e}} = \int \frac{dm_{B}}{m_{u}} l_{\bar{\nu}_{e}}, 
\end{equation}
where $m_{B}$ is the baryonic mass coordinate 
and $m_{u}$ is the atomic mass unit.  
Cooling due to neutrino emission enters in the 
energy equation and entropy equation.  
The cooling term $Q_{cooling}$ is expressed as, 
\begin{equation}
Q_{cooling} = l_{\bar{\nu}_{e}}.  
\end{equation}

\subsection{Nuclear Fission}
During the expansion,
because of neutron captures followed by $\beta$-decays, 
the proton and neutron numbers of nuclei increase 
and they may become unstable to spontaneous fission.  
Fission of nuclei is treated in a very simple manner 
in this current study.  We assume that a nucleus fissions  
spontaneously once its proton number $Z$ 
satisfies the condition 
\begin{equation}
Z \geq Z_{0}, 
\end{equation}
where $Z_{0}$ is the proton number which defines the spontaneous
fission line. We take $Z_{0}=90$ in the current study.  
This criterion for fission is rather simple but 
it allows us to study the essential effects caused by this extra heating
on the hydrodynamics. In a forthcoming paper we will incorporate
a more realistic treatment of fission
such as the one used in the study by Colpi et al. (1993) 
(see also Lattimer et al. 1977). 
We assume further that fission is symmetric and set the proton and 
neutron numbers to be one half of those of the parent nuclei.  
We set also the number fraction of nuclei to twice 
the original value.  
We evaluate the liberated energy 
from the difference of the total energy of the matter 
before and after fission.  This energy contributes 
to the heating and enters into the entropy equation.
The heating term due to fission is 
expressed as 
\begin{equation}
Q_{fission} = \frac
{\epsilon (\rho, Y_{e}, 2Y_{A}, A/2, Z/2) 
- \epsilon (\rho, Y_{e}, Y_{A}, A, Z)} {n_{B}}, 
\end{equation}
where $\epsilon (\rho, Y_{e}, Y_{A}, A, Z)$ is 
the total energy density of the matter at the density 
$\rho$ with the composition $Y_{e}$, $Y_{A}$, $A$, and $Z$.  

\subsection{The Initial Model}\label{chap:ini}
An initial model is constructed as follows. 
We solve the Tolman-Oppenheimer-Volkoff equation with
the Harrison-Wheeler equation of state 
to obtain a series of neutron star models in hydrostatic
equilibrium and at zero temperature.  
For the configuration with the lowest mass
we find $0.189 M_{\odot}$.  
The central density of this neutron star is $2.67 \times 10^{13} 
g/cm^{3}$ and its radius is $230 km$.  
% cf Colpi Non Rel
As an initial perturbation 
we remove a certain amount of mass from the surface of 
the hydrostatic configuration,
given by the mass $\Delta M$ removed in units of the solar mass:
\begin{equation}
\delta \equiv \frac{\Delta M}{M_{\odot}}.
\end{equation}
Removing a surface layer from the star physically means 
that mass is lost on the dynamical time scale.   
This implies a certain scenario for the onset of the instability
and may not be realistic. But in order to be able to compare 
with  previous works  we use instantaneous mass loss as an initial
perturbation here.

We checked that the initial model remained hydrostatic if
no perturbation 
(i.e. $\delta$ = 0) was imposed.
For practical reasons, we set the initial temperature in the 
whole star to $10^{8} K$ when we map the hydrostatic 
configuration onto the hydrodynamics code.  
This temperature is sufficiently low such that it does not 
change the hydrostatic equilibrium because also 
the equation of state is essentially unchanged.   

\section{Numerical Results}
We have carried out numerical simulations of the evolution of 
neutron stars below the minimum mass by changing the value 
of the mass-loss parameter $\delta$.  In the present paper, 
we report the results obtained for one particular choice
of $\delta$ and concentrate on 
the question of the star's final fate, namely whether it explodes or not.  
The details of the results for various values 
of $\delta$ will be given elsewhere.  

A similar study of the neutron stars below the minimum mass 
has been carried out by Colpi et al. (1993) and, in fact, we have chosen  
micro-physics input and initial model in such a way to be able
to compare our results with their's directly.
They found that a large amount of  mass, more than
$\delta=0.20$, had to be removed to initiate a dynamical 
instability, which then lead to the disruption of the whole star. 
However, they had to assume  either
$\beta$-equilibrium  
or had to  artificially increase the $\beta$-decay rates   
in order to enter into the dynamical phase.
Moreover, they could follow the evolution only over
about 0.1 second, and the final fate could not be clarified 
for more realistic $\beta$-decay rates. In particular,
in their paper Colpi et al. (1993) calculated the fate of
a neutron star with normal $\beta$-decay rates and 
$\delta=0.22$ up to 0.1 second and found that 
the outer layers  of the star were expanding, but the 
inner parts were still near to hydrostatic equilibrium.
They inferred that an explosion 
should result on a time scale of about 100 seconds by extrapolating 
the results they had obtained for unrealistically fast $\beta$-decay rates.  
Here, we report the results of numerical simulations which follow the 
evolution over sufficiently long times avoiding artificial speed-ups 
of $\beta$-decays, and we chose also
$\delta=0.22$ as in Colpi et al. (1993) in order to be consistent 
with their work.

\subsection{A Delayed Explosion}
We show in Fig. 1 the results of our numerical simulations 
for $\delta=0.22$.  The position of  selected mass-zones
is plotted as a function of time. Shown are the trajectories 
of every 5th zone.  The outer parts start expanding after the 
mass has been removed because they are no longer in pressure equilibrium. 
In addition, these shells move out of $\beta$-equilibrium 
because the density decreases quickly. Subsequent $\beta$-decays 
heat the matter and increase the electron fraction, leading to 
an increase of the pressure.  This feedback accelerates the matter 
further and more shells begin to expand. The net result is a leptonization 
wave progressing inwards and affecting larger and larger fractions
of the star. However, although
the expansion of the outer parts 
continues due to this coupling of hydrodynamics and 
$\beta$-decays and  the expansion becomes even faster
as time proceeds due to shorter $\beta$-decays times,
the central core stays almost hydrostatic 
for about 20 seconds.  The region of expansion grows slowly 
inwards, nuclei form in the core region because of the decreasing density
and undergo $\beta$-decays,
and finally after 21 seconds also the core of 
the star becomes unstable and explodes.  

\begin{figure}
%\picplace{8cm}
\centerline{\psfig{figure=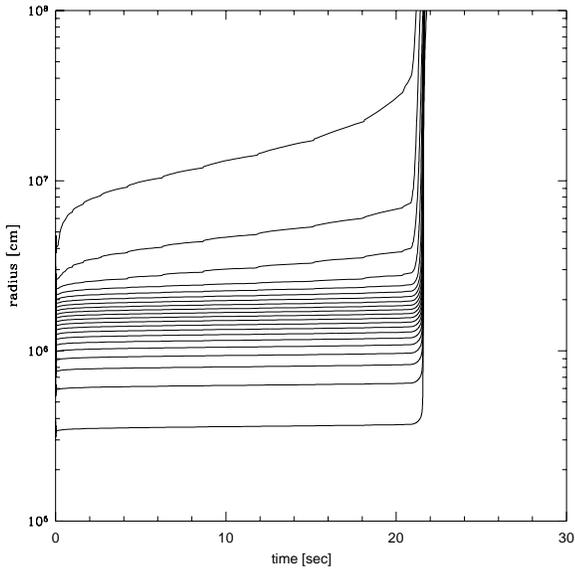,width=8cm,height=8cm}}
\caption[]{The position of selected mass-zones of 
           the neutron star 
           is plotted as a function of time.
           The trajectories of every 5th mass-zone 
           are shown.  
%          After 21 seconds, the whole 
%          star becomes unstable and explodes. (See Fig. 2)
          }
\end{figure}

\begin{figure}
%\picplace{8cm}
\centerline{\psfig{figure=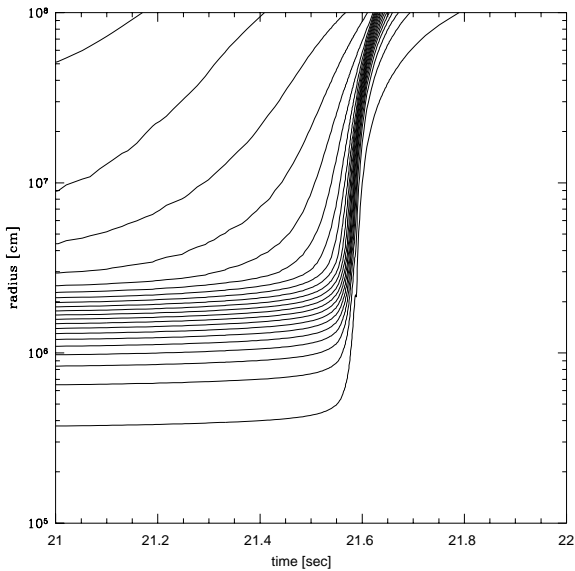,width=8cm,height=8cm}}
\caption[]{The position of selected mass-zones 
           is plotted as in Fig. 1, but expanded for the time 
           between 21 and 22 second.
          }
\end{figure}

This final explosion of the core is very violent and takes place 
on a short time scale of 0.1 seconds only.  We show in Fig. 2 
the trajectories of mass elements as in Fig. 1, but expanded for 
the time between 21 and 22 second after the onset of the calculations.
This part of the explosion  
looks similar to the result obtained  
by Colpi et al. (1993) for much faster (and unrealistic)
$\beta$-decay rates. So 
we conclude that the neutron star explodes in any case, and 
that $\beta$-decays are crucial
for the onset of the explosion but matter less at later times.
 
In summary, we arrive at a first and important conclusion:
Neutron stars near the minimum mass do explode if a moderate
amount of mass is peeled off from their surface on 
a dynamical time scale. The rather poorly known $\beta$-decay rates
of neutron-rich nuclei strongly affect the duration of
the first quasi-static expansion phase but not so much
the final outcome. The energy of the explosion comes mainly
from nuclear decay energy, both from $\beta$-decays and from fission,
and neutrino losses cannot be ignored but are not crucial. 
Thermodynamically, the matter of the expanding neutron star
stays sufficiently close to a local $\beta$-equilibrium
during the early phase and, therefore, the effective adiabatic index 
stays low, below 4/3, and the star does not find a new equilibrium  
configuration. Moreover, once the core region starts to expand,
because of the low initial lepton fraction the nuclear energy release is 
big and fast enough to give rise to a violent explosion.
 
\subsection{The Thermal History of the Expanding Neutron Star Matter
and Nucleosynthesis}
Next we discuss
the thermal history of the expanding  neutron star matter 
during the explosion. This is primarily interesting because it 
determines the nucleosynthesis predictions.
We display the main results
in Fig. 3 as trajectories of mass shells in the 
temperature - density plane.  We show trajectories only
of mass elements which are part of the central regions
of the star. They correspond to one half of the total mass and 
every 5th mass-zone is given. 
It is remarkable that all trajectories are bunched 
together and that the thermal history is quite similar for 
all of them.  
We note that the thermal history of the rest of the 
star is also similar, but with a somewhat broader range around 
the ones shown in Fig. 3.  

\begin{figure}
%\picplace{8cm}
\centerline{\psfig{figure=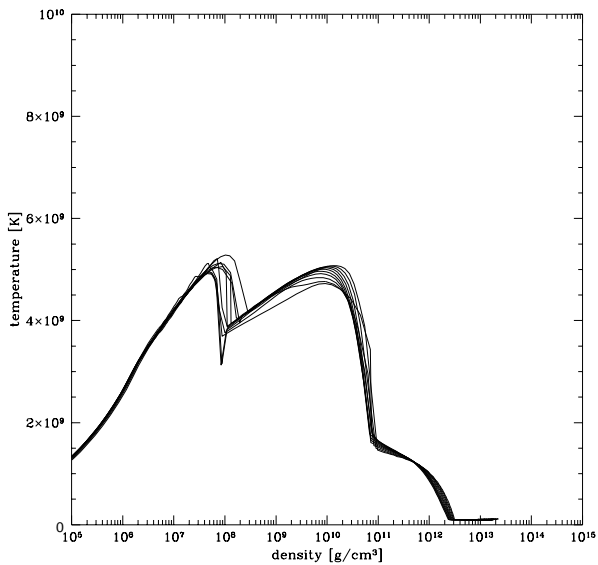,width=8cm,height=8cm}}  
\caption[]{Thermal history of neutron star matter during the explosion 
           is displayed in the temperature - density plane.  
           Shown are the trajectories of mass-zones (one half of the total 
           mass and every 5th zone) in the central region of 
           the star.  
          }
\end{figure}

The trajectories all start from a density above $10^{13} g/cm^{3}$ 
at the initial temperature of $10^{8} K$, and the density decreases due to 
the expansion.  The temperature stays constant down to the transition 
density $\rho_{tr}$ since no nuclear reaction take 
place in the core and, for practical reasons, we do not allow the temperature
to drop below 10$^8 K$.
(This would be different for a more realistic 
equation of state.)
The matter is converted into nuclei with dripped 
neutrons at $\rho_{tr}$ and, from then on,
is heated up due to $\beta$-decays despite the fact that it is
expanding. It can clearly be seen in Fig. 3
that $\beta$-decay heating dominates over adiabatic losses.
A large jump of the temperature at
a density near $10^{11} g/cm^{3}$ 
indicates the onset of fission of nuclei and the temperature reaches  
peak values close to
$5 \times 10^{9} K$.  From then on rapid expansion leads
to some cooling ($\beta$-decay heating can no longer compete
with adiabatic losses) until
once again fission sets in and heats the matter once more
to temperatures near $5 \times 10^{9} K$, and finally the expansion 
cools down the matter continuously.

The fact that rather
high temperatures are reached during the explosion is important
for the nucleosynthesis yields expected from such an event.  
On the one hand side,
the peak temperatures are considerably higher than $10^{9} K$, so  
significant processing by
charged particle nuclear reactions can proceed.  
On the other hand side, the temperature is not too high,  
e.g. lower than about
$5 \times 10^{9} K$, so that a nuclear statistical equilibrium, 
which would erases the memory of the original composition, 
is never fully achieved.  Therefore, nucleosynthesis 
proceeds with various nuclear reactions, but keeping some 
memory of neutron-rich cold neutron star matter.  

Cold expansion of neutron star matter has been discussed as
a possible site for r-process nucleosynthesis,
and the abundances of r-elements produced in the expansion 
have been calculated by parameterizing plausible scenarios
for the dynamics of the star 
(Sato 1974; Lattimer et al. 1977; Meyer 1989).  
The crucial questions concerning r-process-like nucleosynthesis
in exploding low-mass neutron stars are whether or not adiabatic
expansion limits $\beta$-decay heating (a question which
has been answered in a positive way in this section) and
whether or not a sufficient number of neutron captures
and $\beta$-decays can produce also nuclei in, say, the
Uranium region. Again, the answer to this question depends on the 
details of the dynamical evolution and, therefore, requires the
use of numerical simulations. As was stated earlier, fission heating
is an important ingredient in this respect, indicating that indeed 
in our computations nucleosynthesis proceeds all the way up to
the spontaneous fission line, and we
found by coupling the hydrodynamics  to the various
nuclear processes
that r-process nucleosynthesis can in fact 
take place during the explosion of a neutron star 
at the minimum mass.
 
\begin{figure}
%\picplace{8cm}
\centerline{\psfig{figure=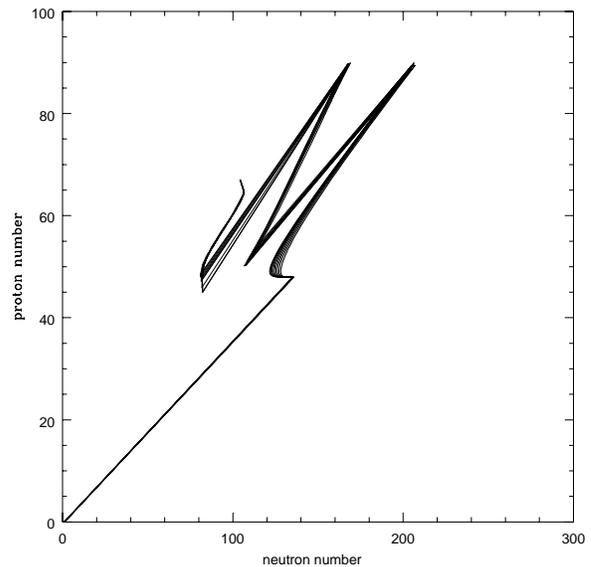,width=8cm,height=8cm}}  
\caption[]{The history of nuclei in neutron star matter 
           during the explosion is displayed in the nuclear chart.  
           Shown are the trajectories of mass-zones 
           in the central region of the star as in Fig. 3.  
          }
\end{figure}

To be a bit more quantitative,
Figure 4 shows again the history of the matter during 
the explosion, but now in the chart of nuclides.
We display the trajectories of the mass elements 
of the core as in Fig. 3.  
It is obvious that the mean nucleus present in the
expanding neutron star matter is very neutron-rich,
neutron-to-proton ratios of up to 2 are realized in
all mass-shells originally belonging to the stellar core,
conditions that are favorable for r-process
nucleosynthesis.
At the beginning of the evolution, 
this matter is composed of neutrons 
with a small fraction of protons and electrons 
and is converted into nuclei with many dripped neutrons 
and some electrons once the density drops below $\rho_{tr}$.  
During the expansion nuclei absorb the dripped neutrons rapidly,
followed by slow
$\beta$-decays, and the mass-number of the mean nucleus
grows towards  heavier elements.  
When the proton number of nuclei reaches $Z_{0}=90$  
(where we have put the spontaneous fission line)
fission takes place and the nuclei split in half, 
according to our simplified treatment.
The light fission products again absorb neutrons rapidly,
with subsequent $\beta$-
decays, until they fission again.  
After two cycles the matter density has become 
so low and the expansion rate is so fast that 
$\beta$-decays do no longer happen and dripped  
neutrons are no longer around and,
as the final product after freeze-out, we find nuclei in
the rare earth region.   
 
This result, although very promising, is not fully conclusive.
Firstly, one has to bear in mind that some of the micro-physics
input used here needs refinements, such as the equation of state,
the treatment of fission, etc. Moreover, we have always
assumed that the matter composition is well characterized
by one representative nucleus. While for calculating
the equation of state this is a rather good approximation,
it is less obvious that it also remains true if we 
are concerned with $\beta$-decay rates
and fission properties. It is interesting to note, however,  
that the various mass-shells follow almost the same
trajectories in the (Z,N)-plane which means that even
with more elaborate input physics the final composition
will be characterized by a typical set of thermodynamic
parameters. This will allow us to post-process 
typical shells with an extended r-process network,
to compute detailed abundances, and to compare them 
with observations.
These aspects of our work are now under investigation.  
Finally, the peak temperatures found in our
computations (close to $5 \times 10^9 K$) may be
too high to ignore charged particle reactions altogether.
If it should turn out that with more realistic equations 
of state we still get peak temperatures of that order
we would have to calculate the nuclear abundances from a 
full reaction network.

\subsection{Neutrino Emission from an Exploding Neutron Star}
We conclude this section with a brief discussion
of the neutrino signal that is to be expected from an
exploding neutron star. In fact, due to the $\beta$-decays
of nuclei
copious neutrinos are emitted, both during
the quasi-static expansion of the neutron star 
and in its final explosion.  
Figure 5 shows the luminosity in electron anti-neutrinos 
as a function of time predicted from our computations.  
The red-shift factor is taken into account, 
but its effect is quite small.  
Typically, we find neutrino luminosities below
$10^{50} erg/s$ during the slow expansion phase
of 20 seconds. The spiky nature of the luminosity
reflects always the onset of fission in some mass-zones, 
because fission not only provides heating of the matter but  
also shorter $\beta$-decay lifetimes than of the parent  
nuclei. Therefore, the neutrino emission suddenly     
increases whenever fission occurs. 

\begin{figure}
%\picplace{8cm}
\centerline{\psfig{figure=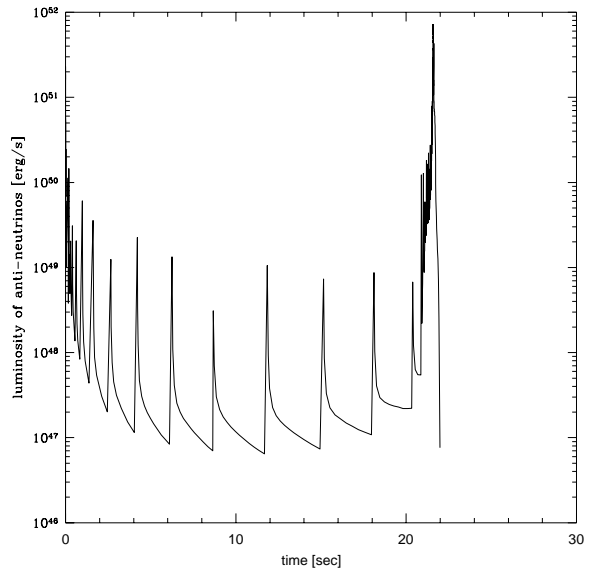,width=8cm,height=8cm}}  
\caption[]{Predicted luminosity of electron anti-neutrinos
           during the evolution of the neutron star 
           is plotted as a function of time.  
          }
\end{figure}

The final explosion is accompanied by a short outburst
of neutrinos lasting for about 0.2 seconds with
a peak luminosity of nearly $10^{52} erg/s$. 
This is still only of the order of 10$\%$ of
the electron (anti-) neutrino luminosity of a core-collapse 
supernova, but it would be observable with existing neutrino
detectors if such an event
occurred in our own Milky Way Galaxy. 
%\subsection{Reconfiguration}

\section{Summary and Conclusions}
We have demonstrated that a neutron star which happens to be close
to the minimum stable mass will explode if of the order of 20$\%$
of its mass are removed from its surface. We found that because
of the reduced gravity the outer layers expand first on
time scales of about 20$s$, leaving the core essentially untouched. 
However, due to slow $\beta$-decays also the core is heated somewhat
and expands. Nuclear energy release, both from $\beta$-decays and
from fission, finally unbinds the core and causes it (and the entire
star) to explode on a time scale of a fraction of a second.
 
The peak temperatures we find never exceed $5 \times 10^9 K$.
This is an interesting and important result because it indicates
that a kind of r-process will take place in exploding neutron stars. 
The abundance yields still have to be calculated, but from this
preliminary study it is already apparent that the temperatures
necessary for a complete nuclear statistical equilibrium are not
reached during the explosion. Therefore, some information from
the pre-explosion phase is preserved and we expect that a significant
fraction of the neutron star's mass is ejected in form of neutron-rich
isotopes of heavy nuclei. This will allow us to place strong constraints
on the event rate of such explosions, no matter whether they result from merging
binary neutron star systems or from fragmenting cores of rapidly
spinning collapsing massive stars. For example, if a total of
$0.1~M_{\odot}$ were ejected with typical solar system r-process
composition (which seems possible on the basis of our present
study) this would imply that the event rate in our Galaxy
should not exceed about 1 per
10$^4$ years to avoid overproduction of r-isotopes. Of course, this
constraint will become more solid once we have performed full
nuclear network calculations on the basis of our hydrodynamical
models.
 
As we have discussed earlier a major short-coming of our study as well
as of all previous ones is the rather unrealistic equation of state
that has been applied. With a more realistic equation of state one
expects (see, e.g., Baym et al. 1971; Colpi et al. 1993) 
that at the minimum mass almost the
entire star will consist of crust material, maybe with a small central
core of a few percent of a solar mass. According to what we discussed in  
Section 2 this implies that even if only a very small amount of mass
is removed from the surface $\beta$-decays will set in everywhere
in the star, leading to faster heating and, likely, to a more
violent explosion. Whether or not the temperatures remain
as low as was found here has to be seen.
 
A second question which has to be addressed in a forthcoming work is
the sensitivity of the predictions to the rather uncertain $\beta$-decay
rates. Here, we have used rates which  are probably too fast 
(see, e.g., Takahashi et al. 1973; Kodama and Takahashi 1975).  
Slower ones will postpone the explosion and counteract the effect of
an equation of state with nuclei up to nuclear saturation density.
But in general terms we expect an effect on the explosion time scale only
but not so much a change of the thermodynamical properties of the
exploding matter. So the net-effect of changing the $\beta$-decay rates
on the nucleosynthesis yields is expected to be small.
 
Finally, we have shown that exploding neutron stars are sources of 
electron anti-neutrinos and can, in principle, be detected by already
existing neutrino detectors if such explosions would occur in our
own Milky Way Galaxy. However, because of the very low event rate discussed
above, this is very unlikely. More distant events, on the other hand, 
would not lead to a measurable signal in present neutrino experiments. 
Because of the compactness of the star the explosion will, however,
give rise to emission of short wave-length electromagnetic radiation. A rough estimate shows 
that this emission will peak in the keV range with luminosities of the
order of up to $10^{47} erg/s$ and a typical duration of around tens of  
seconds. Of course, despite of this enormous X-ray luminosity which in
principle would allow to observe neutron star explosions at large distances,
in practice, however, because of the low event rate and the short duration of the burst,
real detections seem not to be very probable.

\begin{acknowledgements}
K. S. is grateful for the support by the Alexander von Humboldt Stiftung 
and the hospitality of Max-Planck-Institut f\"ur Astrophysik.  
We wish to express our thanks to K. Takahashi and T.-H. Janka 
for many comments and useful suggestions concerning that physics and numerics.  
This work is partially supported by Japan Society for the Promotion of 
Science (JSPS) Postdoctoral Fellowships for Research Abroad.
It was also in part supported by the Sonderforschungsbereich
SFB 375 "Astro-Teilchenphysik" of the Deutsche Forschungsgemeinschaft.
\end{acknowledgements}


\begin{thebibliography}{}
\bibitem{}
Baym, G., Pethick, C.\ J., Sutherland, P.\ G., 
1971, ApJ 170, 299

\bibitem{}
Blinnikov, S.\ I., Novikov, I.\ D., 
Perevodchikova, T.\ V., Polnarev A.\ G., 
1984, Sov.\ Astron.\ Lett. 10, 177

\bibitem{}
Clark, J.\ P.\ A., Eardley, D.\ M., 
1977, ApJ 215, 311

\bibitem{}
Colpi, M., Shapiro, S.\ L., Teukolsky, S.\ A., 
1989, ApJ 339, 318

\bibitem{}
Colpi, M., Shapiro, S.\ L., Teukolsky, S.\ A., 
1991, ApJ 369, 422

\bibitem{}
Colpi, M., Shapiro, S.\ L., Teukolsky, S.\ A., 
1993, ApJ 414, 717

\bibitem{}
Eichler, D., Livio, M., Piran, T., 
Schramm D.\ N., 
1989, Nat 340, 126

\bibitem{}
Harrison, B.\ K., Thorne, K.\ S., Wakano, M., 
Wheeler, J.\ A., 
1965, Gravitation Theory and Gravitational Collapse, 
University of Chicago Press, Chicago

\bibitem{}
Imshennik, V.\ S., Popov, D.\ V., 
1996, MPA preprint, No. 940, submitted to A\&A

\bibitem{}
Kodama, T., Takahashi, K.,
1975, Nucl. Phys. A239, 489

\bibitem{}
Lai, D., Rasio, F.\ A., Shapiro, S.\ L.,
1994, ApJ 420, 811

\bibitem{}
Lattimer, J.\ M., Mackie, F., Ravenhall, D.\ G., 
Schramm, D.\ N., 
1977, ApJ 213, 225

\bibitem{}
Lattimer, J.\ M., Pethick, C.\ J., Ravenhall, D.\ G., 
Lamb, D.\ Q., 
1985, Nucl. Phys. A432, 646

\bibitem{}
Lyne, A.\ G., Lorimer, D.\ R.,
1994, Nat 369, 127

\bibitem{}
Meyer, B.\ S., 
1989, ApJ 343, 276

\bibitem{}
Ruffert, M., Janka, H.-T., Schaefer, G.,
1996, A\&A 311, 532

\bibitem{}
Ruffert, M., Janka, H.-T., Takahashi, K., 
Schaefer, G.,
1997, A\&A 319, 122

\bibitem{}
Sato, K.
1974, Prog. Theor. Phys. 51, 726

\bibitem{}
Shapiro, S.\ L., Teukolsky, S.\ A.,
1983, Black Holes, White Dwarfs, and Neutron Stars: 
The Physics of Compact Objects, Wiley, New York

\bibitem{}
Takahashi, K., Yamada, M., Kondoh, T.,
1973, Atomic Data and Nucl. Data Tables 12, 101

\bibitem{}
Yamada, S., 
1997, ApJ 475, 720

\end{thebibliography}
\end{document}